\documentclass{emulateapj}

\usepackage{graphicx}
\usepackage{float}
\usepackage{amsmath}
\usepackage{epsfig,floatflt}

\begin{document}

\title{An estimate of the primordial non-Gaussianity parameter \MakeLowercase{f}$_{NL}$ using the needlet bispectrum from WMAP}

\author{
\O{}ystein Rudjord\altaffilmark{1,2},
    Frode K. Hansen\altaffilmark{2},
    Xiaohong Lan\altaffilmark{3},\\
    Michele Liguori\altaffilmark{4},
    Domenico Marinucci\altaffilmark{3},
    Sabino Matarrese\altaffilmark{5}}

\altaffiltext{1}{email: oystein.rudjord@astro.uio.no}
\altaffiltext{2}{Institute of Theoretical Astrophysics, University
of Oslo, P.O.\ Box 1029 Blindern, N-0315 Oslo, Norway}
\altaffiltext{3}{Dipartimento di Matematica, Universit\`a di Roma
`Tor Vergata', Via della Ricerca Scientifica 1, I-00133 Roma, Italy}
\altaffiltext{4}{Department of Applied Mathematics and Theoretical
Physics, Centre for Mathematical Sciences, University of Cambridge,
Wilberfoce Road, Cambridge, CB3 0WA, United Kingdom}
\altaffiltext{5}{Dipartimento di Fisica, G. Galilei, Universit\`a di
Padova and INFN, Sezione di Padova,via Marzolo 8,I-35131 Padova,
Italy}


\begin{abstract}
We use the full bispectrum of spherical needlets applied to the WMAP
data of the cosmic microwave background as an estimator for the
primordial non-Gaussianity parameter $f_{NL}$. We use needlet scales up to $\ell_\mathrm{max}=1000$ and the KQ75 galactic cut and find $f_{NL}=84\pm40$ corrected for point source bias. We also introduce a set of consistency tests to validate our results against the possible influence of foreground residuals
or systematic errors. In particular, fluctuations in the value of
$f_{NL}$ obtained from different frequency channels, different masks
and different multipoles are tested against simulated maps. All
variations in $f_{NL}$ estimates are found statistically consistent
with simulations.
\end{abstract}


\section{Introduction}
\label{sec:introduction}
The theory of inflation predicts the fluctuations in the Cosmic
Microwave Background (CMB) to be close to Gaussian
distributed. However, a small degree of non-Gaussianity is generally
present in all the inflationary scenarios. The primordial non-Gaussian
signal predicted by many models can be parametrized in the form:
\begin{equation}\label{eqn:primordialNG}
\Phi(\mathbf{x}) = \Phi_L(\mathbf{x}) + f_{\rm NL}
\left(\Phi_L^2(\mathbf{x}) - \langle \Phi_L^2(\mathbf{x}) \rangle \right) \; ,
\end{equation}
where $\Phi(\mathbf{x})$ is the primordial curvature perturbation
field at the end of inflation and $\Phi_L(\mathbf{x})$ is the Gaussian
part of the perturbation. The dimensionless parameter $f_{\rm NL}$
describes the amplitude of non-Gaussianity.
The non-Gaussian part of the primordial curvature perturbation is a local functional
of the Gaussian part and for this reason this kind of parametrization
is often referred to as {\em local} non-Gaussianity. Local
non-Gaussianity is predicted to arise from standard single-field
slow-roll inflation \citep{standard1,standard2} as well as from alternative inflationary
scenarios for the generation of primordial perturbations, like
the curvaton \citep{curvaton1,curvaton2,curvaton3} or inhomogeneous (pre)reheating models \citep{gamma1,gamma2,gamma3}, or even from
alternatives to inflation, such as ekpyrotic and cyclic models
\citep{ek1,ek2}. Other models, such as DBI inflation \citep{dbi} and ghost
inflation \citep{ghost},
predict a different kind of primordial non-Gaussianity, called
  "equilateral", because the three point function for this kind of
non-Gaussianity is peaked on equilateral configurations, in which the
lengths of the three wavevectors forming a triangle in Fourier space are
equal \citep{creminelli}. In this paper we will focus only on
non-Gaussianity of the local type, described by equation
\ref{eqn:primordialNG}. The interesting aspect of primordial
non-Gaussianity is that the expected non-Gaussian amplitude $f_{\rm
  NL}$ varies significantly from model to model. Putting
experimental bounds on $f_{\rm NL}$ is then equivalent to constraining
primordial scenarios of inflation. For example standard single-field
slow-roll inflation predicts $f_{\rm NL} \sim
10^{-2}$ at the end of inflation \citep{standard1,standard2}  (and therefore a final value
$\sim $ unity  after general relativistic second-order perturbation
effects are taken into account \citep{second1,second2}). Such a small value is
not experimentally detectable and for this reason an eventual
detection of a Gaussian signal in present and forthcoming CMB data
will rule out single-field slow-roll inflation as a viable
scenario. Motivated by these considerations many groups have attempted
to measure  $f_{\rm NL}$ using CMB datasets, and WMAP data in
particular.

A detection of non-zero $f_{NL}$ at more than the $2\sigma$ level was
found by \citep{amit} using the WMAP data with the Kp0 galactic cut.
The WMAP team found similar values but stating that only the value obtained
with the slightly larger KQ75 galactic cut is reliable due to possible foreground residuals.
In this case a value of $f_{NL}=51\pm32$ was found. In both these cases,
an extended version \citep{creminelli,amit,amit3} of the \citep{ksw}
(KSW-method) based on the full bispectrum was used. Consistent results were found
by \citep{curto} and \citep{pietrobon} using parts of the bispectrum of
spherical mexican hat wavelets \citep{smhw} and the skewness of needlet coefficients. A recent estimate has now been made by \citep{smith} obtaining the smallest error bars on $f_{NL}$ so far finding $f_{NL}=38\pm21$.

Recently, it was shown in \citep{needbisp} that needlet coefficients
can be used to construct statistics which are directly related to
the bispectrum. These statistics share most of the useful
properties of the bispectrum, while at the same time they do present
important advantages, especially in terms of robustness to masked
data and computational rapidity. Motivated from these results, in
this paper we will use the full bispectrum of needlet coefficients
as an estimator of $f_{NL}$, introducing moreover a set of
consistency tests to check the stability of our findings. In
particular, we shall investigate whether changes in the estimated
values of $f_{NL}$ using different galactic cuts, different
frequency bands and different multipoles are within the variations
expected from simulations. Of course, variations in $f_{NL}$ among
these different cases which are significantly larger than
statistical fluctuations might point out the presence of foreground
residuals or other systematic effects that could have biased the
estimate of $f_{NL}$. 

The plan of the paper is as follows. In section \ref{sec:data}, we describe the data used in the analysis. Then in section \ref{sec:method}, the needlets and the needlet bispectrum estimator are described in detail. Finally, the results on the WMAP data are presented in section \ref{sec:results} and conclusions are made in section \ref{sec:concl}.

 \section{Data}
\label{sec:data}
For this analysis we used the noise weighted average of the V and W
frequency band of the WMAP 5 year CMB map, as well as the corresponding instrumental
beam and noise properties. We have also performed the analysis on the
individual Q (41 GHz), V (61 GHz) and W (94 GHz) bands.  For masking out galactic foregrounds and
point sources, we used the $KQ75$ and $KQ85$ mask supplied by the WMAP
team. For particular cases, we also used the much smaller Kp12 mask (maintaining $94\%$ of the sky) as well as an extended KQ75+ mask. The KQ75+ mask is constructed from the KQ75 mask, extending the mask with 5 degrees along the rim, maintaining a total of $63\%$ of the sky. We have used the maps at
Healpix\footnote{http://healpix.jpl.nasa.gov} resolution $N_{side} = 512$.

\section{Method}
\label{sec:method}

\subsection{Spherical needlets}

Needlets are a new form of (second generation) spherical wavelets,
which were introduced into functional analysis by (\citep{npw1,npw2})
and have attracted a lot of attention in the cosmological literature
hereafter. The possibility to use needlets for the statistical
analysis of spherical random fields, with a view to CMB
applications, is first discussed in \citep{bkmpAoS}, where the
stochastic properties of needlet coefficients are established and
their possible roles for data analysis (spectrum estimation,
Gaussianity testing) are described; further mathematical properties
where then given in \citep{bkmpBer}. The first application to CMB
data, in particular, for the analysis of cross-correlation of CMB
and Large Scale Structure data was provided by \citep{pbm06}; a
general presentation of the method for a CMB audience is given in
\citep{mpbb08}, while in \citep{guilloux} the properties of different
weighting schemes are investigated and compared. Further papers have
applied needlets on CMB data, for issues such as map-making,
spectrum estimation, detection of features and anisotropies
(\citep{fg08,dela08,fay08, pietrobon08}); more recently, needlets
have also been considered for the analysis of directional data, with
a view to high energy cosmic rays (\citep{denest}) and for the
analysis of polarization data (\citep{spin-mat,ghmkp08}), whereas
extensions to the so-called Mexican needlets case are discussed by
\citep{gm1,gm2}, their stochastic properties being established
in\citep{lan2,Mayeli}.

The spherical needlet (function) is defined as
\begin{equation}
\psi _{jk}(\hat{\gamma})=\sqrt{\lambda _{jk}}\sum_{\ell }b(\frac{\ell }{B^{j}%
})\sum_{m=-\ell }^{\ell }\overline{Y}_{\ell m}(\hat{\gamma})Y_{\ell
m}(\gamma _k)\text{ };  \label{needlets_expansion}
\end{equation}%
here, $\hat{\gamma}$ is a direction on the sphere, and $j$ is the
frequency (multipole range) of the needlet and  $\lambda _{jk}$ is a
normalizing factor. The points $\left\{ \gamma
_k\right\} $ can be identified with the pixel centres in the HealPix
pixelization scheme. The number B defines the needlet basis such that only multipoles in the range $\ell=[B^{j-1},B^{j+1}]$ are included, i.e. the function $b(\ell/B^j)$ is zero outside this range. For details in the functional form of $b(\ell/B^j)$. please refer to \citep{mpbb08} and references therein.

The advantages of needlets have already been discussed in several
papers in the literature; in short, we recall that needlets do not
rely on any tangent plane approximation; they are computationally
very convenient, and inherently adapted to standard packages such as
HealPix; they allow for a simple reconstruction formula (a property
which is \emph{not }shared by other wavelets systems); they are
quasi-exponentially (i.e. faster than any polynomial) concentrated
in pixel space. 
Moreover, needlets are exactly localized on a finite number of
multipoles (the width of this support is explicitly known and can be
specified as an input parameter, see Eq.~\ref{needlets_expansion})).

Random needlet coefficients can be shown to be asymptotically uncorrelated (and
hence, in the Gaussian case, independent) at any fixed angular
distance, when the frequency increases. This capital property is in
general not shared by other wavelet systems (see
(\citep{lan2,Mayeli})) and can be exploited in several statistical
procedures, as it allows one to treat needlet coefficients as a
sample of independent and identically distributed coefficients on
small scales, at least under the Gaussianity assumption.

In this paper, for notational simplicity we shall take random
needlet coefficients to be
\[\beta_{jk}=\sum_{\ell }b(\frac{\ell }{B^{j}})\sum_{m=-\ell }^{\ell
}a_{\ell m}Y_{\ell m}(\widehat{\gamma }_{k})\equiv\sum_{\ell m} b_\ell a_{\ell m} Y_{\ell m}(\widehat{\gamma }_{k}) \]

Here $j$ denotes the frequency of the coefficient and
$\widehat{\gamma }_{k}$ denotes the direction on the sky. 
We are dropping a normalizing term
$\sqrt{\lambda _{jk}};$ this comes at no cost, because in this paper
needlet coefficients always appear after normalization with their
own standard deviation, so that this deterministic factor cancels.
From the notational point of view, however, this allows a major
simplification, as it permits to avoid considering different weights
at different frequencies $j.$ As before, the index $k$ can in
practice be the pixel number on the HEALPix grid.

\subsection{The needlets bispectrum}

Starting from some highly influential papers (\citep{Hu,ks}), the
bispectrum has emerged in the last decade as the most promising
statistics for the detection of non-Gaussianity in CMB data. To fix
notation, we recall that, under the
assumption of statistical isotropy for CMB radiation, we must have (\citep{Hu})%
\[
\langle a_{\ell_{1}m_{1}}a_{\ell_{2}m_{2}}a_{\ell_{3}m_{3}}\rangle =\left(
\begin{array}{ccc}
\ell_{1} & \ell_{2} & \ell_{3} \\
m_{1} & m_{2} & m_{3}%
\end{array}%
\right) B_{\ell_{1}\ell_{2}\ell_{3}}\text{ ,}
\]%
where on the right hand side we have introduced the Wigner's 3j
coefficients, which are different from zero only for configurations
of $l_1,l_2,l_3$ which satisfy the triangle conditions (see again
\citep{Hu,ks}). The \textit{unreduced bispectrum
}$B_{\ell_{1}\ell_{2}\ell_{3}}$ is identically zero in
the Gaussian case; under non-Gaussianity, it can be estimated by%
\[
\widehat{B}_{\ell_{1}\ell_{2}\ell_{3}}=\sum_{m_{1}m_{2}m_{3}}\left(
\begin{array}{ccc}
\ell_{1} & \ell_{2} & \ell_{3} \\
m_{1} & m_{2} & m_{3}%
\end{array}%
\right) a_{\ell_{1}m_{1}}a_{\ell_{2}m_{2}}a_{\ell_{3}m_{3}}\text{ .}
\]%
It was shown by (\citep{ks}) that, in the idealistic circumstance
with the absence of masked regions, the bispectrum can constrain
non-Gaussianity very efficiently, with a signal-to-noise ratio equal
to unity for $f_{NL}$ smaller than 10 at the Planck resolution. In
the presence of masked regions, however, these properties
deteriorate consistently; many statistical solutions have been
discussed so far, see for instance \citep{ykwlhm,amit} for the most
recent developments. A large literature has also focussed on the
determinations of the multipole configurations where the
signal-to-noise ratio should be expected to be stronger, in view of
a given model: see for instance \citep{creza,cabella06,bartoloreview,m,amit}
and many others.

Our purpose in this paper is to combine ideas from the bispectrum
and the needlets literature, to propose a needlet bispectrum method
to test non-Gaussianity and estimate the nonlinearity parameter
$f_{NL}.$ More
precisely, we suggest to focus on the needlet bispectrum, defined by%
\[
I_{j_{1}j_{2}j_{3}}=\sum_{k}\frac{\beta_{j_{1}k}\beta_{j_{2}k}\beta_{j_{3}k}}{\sigma
_{j_{1k}}\sigma _{j_{2}k}\sigma _{j_{3}k}}\text{ ,}
\]%
where $\sigma _{j}=\sqrt{<\beta_{jk}^{2}>}$ is the standard
deviation of $\beta_{jk}. $ The needlet bispectrum was first
considered in (\citep{needbisp}), and we refer to that paper for more
discussion and details on its mathematical properties; the use of
needlets for a non-Gaussianity test is also proposed in \citep{pietrobon},
where the focus is instead on the skewness of the coefficients
(which can be viewed as a special case of the bispectrum, for
$j_{1}=j_{2}=j_{3})$. Of course, many other papers had previously
used wavelet-related techniques to search for non-Gaussianity in
CMB, see for instance \citep{vielva04,Cruz1,Cruz2,mcewen2}.

In short, to understand the motivations of our proposals, note that,
denoting by $N$ the cardinality of the points $k$ (i.e., the number
of points in the pixelization scheme, so that $4\pi /N$ provides an
approximation for the pixel area), and neglecting for simplicity
beam
factors, we have approximately%
\[
\frac{4\pi }{N}\sum_{k}\beta_{j_{1}k}\beta_{j_{2}k}\beta_{j_{3}k}=
\]%
\begin{eqnarray*}
&=&\sum_{\ell_{1}m_{1}}\sum_{\ell_{2}m_{2}}\sum_{\ell_{3}m_{3}}%
b(\frac{\ell_{1}}{B^{j_{1}}})b(\frac{\ell_{2}}{B^{j_{2}}})b(\frac{\ell_{3}}{B^{j_{3}}})\\
&\times& \frac{4\pi }{N}\sum_{k}a_{\ell_{1}m_{1}}a_{\ell_{2}m_{2}}a_{\ell_{3}m_{3}}Y_{\ell_{1}m_{1}}(%
\widehat{\gamma }_{k})Y_{\ell_{2}m_{2}}(\widehat{\gamma }_{k})Y_{\ell_{3}m_{3}}(%
\widehat{\gamma }_{k}) \\
\end{eqnarray*}
\begin{eqnarray*}
&\simeq &\sum_{\ell_{i}m_{i}}b(\frac{\ell_{1}}{%
B^{j_{1}}})b(\frac{\ell_{2}}{B^{j_{2}}})b(\frac{\ell_{3}}{B^{j_{3}}}%
)a_{\ell_{1}m_{1}}a_{\ell_{2}m_{2}}a_{\ell_{3}m_{3}}\\
&\times& \int Y_{\ell_{1}m_{1}}(\widehat{%
\gamma })Y_{\ell_{2}m_{2}}(\widehat{\gamma })Y_{\ell_{3}m_{3}}(\widehat{\gamma })d%
\widehat{\gamma }\text{ .}
\end{eqnarray*}%
Write
\[
h_{\ell_{1}\ell_{2}\ell_{3}}=\left(
\begin{array}{ccc}
\ell_{1} & \ell_{2} & \ell_{3} \\
0 & 0 & 0%
\end{array}%
\right) \sqrt{\frac{(2\ell_{1}+1)(2\ell_{2}+1)(2\ell_{3}+1)}{4\pi
}}\text{ ;}
\]%
hence we obtain%
\begin{eqnarray*}
&&\sum_{\ell_{i}m_{i}}b(\frac{\ell_{1}}{B^{j_{1}}%
})b(\frac{\ell_{2}}{B^{j_{2}}})b(\frac{\ell_{3}}{B^{j_{3}}})\\
&\times& a_{\ell_{1}m_{1}}a_{\ell_{2}m_{2}}a_{\ell_{3}m_{3}}\left(
\begin{array}{ccc}
\ell_{1} & \ell_{2} & \ell_{3} \\
m_{1} & m_{2} & m_{3}%
\end{array}%
\right) h_{\ell_{1}\ell_{2}\ell_{3}} \\
\end{eqnarray*}
\begin{eqnarray*}
&=&\sum_{\ell_{1}\ell_{2}\ell_{3}}b(\frac{\ell_{1}}{B^{j_{1}}})b(\frac{\ell_{2}}{B^{j_{2}}}%
)b(\frac{\ell_{3}}{B^{j_{3}}})h_{\ell_{1}\ell_{2}\ell_{3}}\widehat{B}_{\ell_{1}\ell_{2}\ell_{3}}%
\text{ .}
\end{eqnarray*}%
From the previous computations, it should be clear that the needlets
bispectrum can be viewed as a smoothed and normalized form of the
standard bispectrum estimator. As usual with wavelet techniques, the
advantage is that, while the standard bispectrum is known to be
heavily affected by the presence of masked regions, needlet
coefficients are much more robust and consequently the needlet
bispectrum makes up a more reliable statistics even for incomplete
maps. Furthermore, the needlet bispectrum is computationally very
convenient, as it does not require the evaluation of Wigner's 3j
coefficients, which is extremely time-consuming.

From the mathematical point of view, further properties
of the needlets bispectrum are discussed by \citep{needbisp}; in particular, it is
shown that, after normalization, $\widehat{I}_{j_{1}j_{2}j_{3}}$ is
asymptotically Gaussian as the frequency increases.  Furthermore, it
can be shown that (under idealistic experimental circumstances) the
values of the needlet bispectrum are asymptotically independent over
different frequencies, so that chi-square statistics can be suitably
implemented. In \citep{needbisp}, some analytic discussion on the power
properties of the needlet bispectrum for a pure Sachs-Wolfe model
were also provided, showing that its expected valued diverges to
infinity at high frequencies. Although those results were derived in
a mathematical setting and did not take into account many features
of CMB data, the simulations in the present paper show (in a much
more realistic setting) that this procedure does have very
satisfactory power properties in the presence of non-Gaussianity.

\subsection{$f_{NL}$ estimator}
\begin{figure}
  \begin{center}
    \includegraphics[height=6cm, width=8cm]{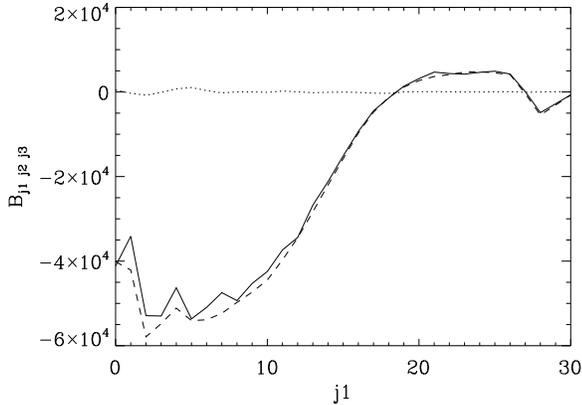}
  \end{center}
  \caption{$\hat I_{j_1 j_2 j_3}$ (dashed) plotted along $j_1 = j$
    while $j_2=25$ and $j_3=25$, compared to the average bispectrum
    from $300$ simulations with $f_{NL}=500$ (full line) and average
    bispectrum from $10000$ Gaussian simulations (dotted).}
  \label{fig:bhat}
\end{figure}

We will now use the needlets bispectrum for estimating $f_{NL}$ by a
$\chi^2$ minimization procedure. We define $\chi^2(f_{NL})$ as
\[
\chi^2(f_{NL})={\bf d}^T(f_{NL}){\bf C}^{-1}{\bf d}(f_{NL}),
\]
where the elements $d_i$ of the data vector $\mathbf{d}$ are defined
as $d_i=I_{j_1j_2j_3}(\mathrm{observed})-\langle
I_{j_1j_2j_3}\rangle(f_{NL})$ for all combinations of
$(j_1,j_2,j_3)$ satisfying the triangle condition. Here
$I_{j_1j_2j_3}(\mathrm{observed})$ is the needlets bispectrum of the
observed data and $\langle I_{j_1j_2j_3}\rangle(f_{NL})$ is the
expectation value of the needlet bispectrum for a given value of
$f_{NL}$. The correlation matrix $\mathbf{C}$ is given by
\[
C_{ij}=\langle d_id_j\rangle-\langle d_i\rangle\langle d_j\rangle.
\]
The correlation matrix is obtained from Gaussian simulations. In
order to avoid cumbersome numeric grid-calculations of $\left<I_{j_1
j_2 j_3}\right>(f_{NL})$, we seek an expression with a more explicit
dependency of $f_{NL}$. In order to arrive at such an expression, we
write out again the needlets bispectrum as
\begin{eqnarray}
&& I_{j_1 j_2 j_3}  =  \sum_k^{npix} \frac{\beta_{{j_1}k}
\beta_{{j_2}k} \beta_{{j_3}k}}
{\sigma_{{j_1}k} \sigma_{{j_2}k} \sigma_{{j_3}k}  } \notag \\
&=& \sum_k \sum_{\ell_1 \ell_2 \ell_3} \sum_{m_1 m_2 m_3}
\frac{b_{\ell_1}}{\sigma_{{j_1}k}}
\frac{b_{\ell_2}}{\sigma_{{j_2}k}}
\frac{b_{\ell_3}}{\sigma_{{j_3}k}}  a_{\ell_1 m_1} a_{\ell_2 m_2}  a_{\ell_3 m_3}  \notag \\
&\times& Y_{\ell_1 m_1}(\widehat{\gamma }_{k}) Y_{\ell_2
m_2}(\widehat{\gamma }_{k}) Y_{\ell_3 m_3}(\widehat{\gamma }_{k})
\label{eq:needlet_bispectrum_def}
\end{eqnarray}
As usual, the non-Gaussian $a_{\ell m}$'s are assumed to be a
combination of a linear (Gaussian) and a non-linear term: $a_{\ell
m} = a_{\ell m}^G + f_{NL}a_{\ell m}^{NG}$. This allows us to write
the three-point correlations in $a_{\ell m}$'s as
\begin{eqnarray}
&& \left<  a_{\ell_1 m_1 }  a_{\ell_2 m_2 }  a_{\ell_3 m_3 }\right>  \notag \\
&=& \left<  a^G_{\ell_1 m_1 }  a^G_{\ell_2 m_2 }  a^G_{\ell_3 m_3 }\right> + f_{NL} \big(
 \left<  a^{NG}_{\ell_1 m_1 }  a^G_{\ell_2 m_2 }  a^G_{\ell_3 m_3 }\right> \notag  \\
&+& \left<  a^G_{\ell_1 m_1 }  a^{NG}_{\ell_2 m_2 }  a^G_{\ell_3 m_3 }\right>
+ \left<  a^G_{\ell_1 m_1 }  a^G_{\ell_2 m_2 }  a^{NG}_{\ell_3 m_3 }\right> \big) \notag \\
&+& \mathcal{O}((a_{\ell m}^{NG})^2)
\end{eqnarray}
The non-linear terms are assumed to be small, and thus we will neglect
the higher order terms, $\mathcal{O}((a_{\ell m}^{NG})^2) \approx
0$. We will also neglect the pure Gaussian term, since the three point
correlation function of a Gaussian field is zero. We insert the
remaining terms into eq. \ref{eq:needlet_bispectrum_def} and find the
mean value:
\begin{eqnarray}
&&\left< I_{j_1 j_2 j_3}\right> (f_{NL})
= f_{NL} \Bigg( \left< \sum_k^{npix} \frac{\beta^{NG}_{{j_1}k}  \beta^G_{{j_2}k}  \beta^G_{{j_3}k}} {\sigma_{{j_1}k} \sigma_{{j_2}k} \sigma_{{j_3}k}  }\right> \notag \\
&+& \left< \sum_k^{npix} \frac{\beta^{G}_{{j_1}k}  \beta^{NG}_{{j_2}k}  \beta^G_{{j_3}k}}{\sigma_{{j_1}k} \sigma_{{j_2}k} \sigma_{{j_3}k}  }\right> \notag \\
&+& \left< \sum_k^{npix} \frac{\beta^{G}_{{j_1}k}  \beta^{G}_{{j_2}k}  \beta^{NG}_{{j_3}k}}{\sigma_{{j_1}k} \sigma_{{j_2}k} \sigma_{{j_3}k}  } \right>\Bigg) \notag \\
&\approx& f_{NL} \left( \left<I^{NG, G, G}_{j_1 j_2 j_3}\right> +
\left<I^{G, NG, G}_{j_1 j_2 j_3}\right>
+ \left<I^{G, G, NG}_{j_1 j_2 j_3}\right>\right) \notag \\
&=& f_{NL}\left<\hat I_{j_1 j_2 j_3}\right>
\end{eqnarray}
where we have defined the quantity
\[
\left<\hat I_{j_1 j_2 j_3}\right> =  \left<I^{NG, G, G}_{j_1 j_2
j_3}\right> + \left<I^{G, NG, G}_{j_1 j_2 j_3}\right> + \left<I^{G,
G, NG}_{j_1 j_2 j_3}\right>
\]
which does not depend on $f_{NL}$ to the first order in $a^{NG}_{\ell
  m}$. Figure \ref{fig:bhat} shows a plot of a average needlet
bispectrum $\langle I_{j_1 j_2 j_3}\rangle(f_{NL})$ from $300$
simulations with $f_{NL}=500$. Here the bispectrum is plotted along
one of the indices $j_1 = j$, while the two other indices $j_2=25$
and $j_3=25$ are kept constant. In the same plot is also our first
order approximation, $500 \times \langle \hat I_{j_1 j_2
j_3}\rangle$. As we see, this approximation is fairly good for
$f_{NL} =500$, and based on previous estimates
(\citep{komatsu:2008}), we will assume that $f_{NL}$ does not have a
value significantly higher than this.

We can now write the elements of the data vector $\mathbf{d}$ of the $\chi^2$ as
 $d_i=I_{j_1j_2j_3}-f_{NL}\langle\hat I_{j_1j_2j_3}\rangle$.
In order to estimate $f_{NL}$, we need to find the value of $f_{NL}$ that minimizes the $\chi^2$
\begin{equation}
\frac{\mathrm{d} \chi^2(f_{NL})}{\mathrm{d} f_{NL}} = 0.
\label{eq:deriv}
\end{equation}
giving
\begin{equation}
  f_{NL} = \frac{\left<\hat I_{j_1 j_2 j_3}\right>^T \mathbf{C}^{-1} I_{j_1 j_2 j_3}(\mathrm{observed})}{\left<\hat I_{j_1 j_2 j_3}\right>^T \mathbf{C}^{-1}\left<\hat I_{j_1 j_2 j_3}\right>}. \label{eq:the_fnl_equation}
\end{equation}

\subsection{The procedure to estimate $f_{NL}$}
\begin{figure}
  \begin{center}
    \includegraphics[height=8cm, width=8cm]{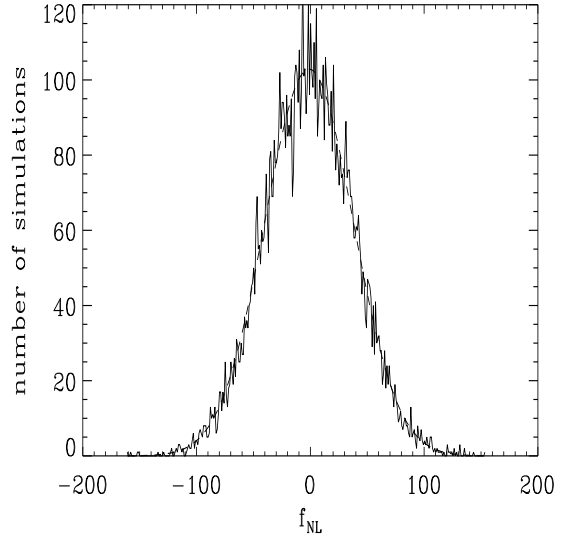}
  \end{center}
  \caption{A histogram (full line) of $f_{NL}$ estimates of $10000$ Gaussian
    simulations of the V+W channel plotted together with a Gaussian fit (dashed). As expected the
    $f_{NL}$ values form an approximate Gaussian distribution with a
    mean value of $f_{NL}=0$.}
  \label{fig:fnl_gaussfit}
\end{figure}

We will now show the full procedure we have used to estimate $f_{NL}$ using the needlet bispectrum.
\begin{enumerate}
\item Generate $10000$ simulations of Gaussian sky maps using the best fit WMAP 5 year power spectrum. These are smoothed with an instrumental beam and noise
  is added. The maps are also multiplied with a mask for galactic cut, in order to remove foregrounds.
  A needlet transform is applied and the standard deviation $\sigma_{jk}$ of the needlet coefficients $\beta_{jk}$ are
  found. This standard deviation is needed to find the needlet bispectrum as seen from
  eq. (\ref{eq:needlet_bispectrum_def}).
\item Produce another $120000$ Gaussian simulations. Mask, beam and noise properties are applied as above.
  The needlet coefficients are found and used to obtain the needlet bispectra, $I_{j_1 j_2 j_3}$.
  These bispectra are used to find the covariance matrix $\mathbf{C}$. This converges very slowly, thus the
  need for such a large number of simulations.
\item Generate $300$ non-Gaussian simulations (\citep{liguori:2007}) to find the mean first order non-Gaussian bispectrum,
  $\langle \hat I_{j_1 j_2 j_3}\rangle$.
\item Obtain the needlet bispectrum from the data and estimate $f_{NL}$ using eq. \ref{eq:the_fnl_equation}
\item Generate a set of 10000 simulated Gaussian maps and estimate $f_{NL}$ in the same manner from these maps in order to obtain the error bars on $f_{NL}$.  This set of estimated
$f_{NL}$ values form a Gaussian distribution around $f_{NL} = 0$.
Figure \ref{fig:fnl_gaussfit} shows a histogram of the $10000$
$f_{NL}$ of the V+W frequency channel estimates plotted together with a Gaussian distribution. We
see that the distribution of $f_{NL}$ estimates is very close to
Gaussian, so we will use the standard deviation as a measure of the
uncertainty of the estimate.
\end{enumerate}

\section{Results}
\label{sec:results}
\subsection{Estimates of $f_{NL}$}

We used the above procedure to estimate $f_{NL}$ from the WMAP data. The co-added V+W map as well as the single frequency bands Q, V and W were used. The estimated values of $f_{NL}$ are
listed in table \ref{table:results} together with the $1 \sigma$
error bars found from the simulations. For the analysis we used both
the $KQ85$ mask and the more aggressive $KQ75$ mask for galactic cut
in order to study potential effects from residual foregrounds.

We have used multipoles up to $\ell_\mathrm{max}=1024$ in the analysis. We tested different values of B in order
to find the number of needlet scales from $\ell=2$ to $\ell=1000$ which
would yield an invertible covariance matrix. We found that the
maximum number of scales which could be used was 31 scales, using
$B=1.2050$. For comparison with the
WMAP results we also used needlet scales including multipoles up to $\ell_\mathrm{max}=500$ and
$\ell_\mathrm{max}=700$. In table \ref{table:results} as well as in the following text, we will use $\ell_\mathrm{max}=1000,700,500$ to specify the highest multipole included in the highest needlet scale. Note that this number may differ slightly from 1000, 700 and 500 depending on the exact value of $B$ specified. For
$\ell_\mathrm{max}=500$, we used 30 scales with $B=1.1828$ and for
$\ell_\mathrm{max}=700$ we used 31 scales with $B=1.1880$.

We see from the table that the best results obtained on the combined V+W band yields $f_{NL}=89\pm39$ for the KQ75 cut and $f_{NL}=117\pm36$ using the smaller KQ85 cut. We run simulations of unresolved point sources based on the procedure described in \citep{komatsu:2008} and obtained a point source bias of $\Delta f_{NL}=5\pm1$ for KQ75 and $\Delta f_{NL}=7\pm1$ for KQ85 giving corrected values of $f_{NL}=84\pm40$ and $f_{NL}=110\pm37$. We see that even for KQ75 a zero value for $f_{NL}$ is excluded at about $2\sigma$. In order to make sure that foregrounds are not influencing our results significantly, we also make an estimate on the much larger KQ75+ mask and obtain $f_{NL}=103\pm41$ or $f_{NL}=97\pm42$ taking into account unresolved point sources.

\begin{table}[htdp]
\begin{center}
\begin{tabular}{|l|c|c|c|c|}
\hline
freq. channel & mask & $\ell_{max}$ & $n_j$ & $f_{NL}$ \\
\hline
V+W & $KQ85$ & $700$ & $31$ & $156\pm 45$ \\
V+W & $KQ75$ & $700$ & $32$ & $88\pm 48$ \\
V+W & $Kp12$ & $1000$ & $31$ & $160 \pm 30$ \\
V+W & $KQ85$ & $1000$ & $31$ & $117\pm 36$ \\
V+W & $KQ75$ & $1000$ & $31$ & $89\pm 39$ \\
V+W & $KQ75+$ & $1000$ & $31$ & $103 \pm 41$ \\
V+W (Raw)& $KQ85$ & $1000$ & $31$ & $105 \pm 36$ \\
V+W (Raw)& $KQ75$ & $1000$ & $31$ & $83 \pm 39$ \\
V+W (Raw)& $KQ75+$ & $1000$ & $31$ & $87 \pm 41$ \\
\hline
V & $KQ75$ & $500$ & $30$ & $78 \pm 57$ \\
V & $KQ85$ & $1000$ & $31$ & $100\pm 39$ \\
V & $KQ75$ & $1000$ & $31$ & $105\pm 42$ \\
V (Raw) & $KQ85$ & $1000$ & $31$ & $88 \pm 39$ \\
V (Raw) & $KQ75$ & $1000$ & $31$ & $100 \pm 42$ \\
\hline
W & $KQ75$ & $500$ & $30$ & $57 \pm 59$ \\
W & $KQ85$ & $1000$ & $31$ & $79 \pm 42$ \\
W & $KQ75$ & $1000$ & $31$ & $54 \pm 45$ \\
W (Raw) & $KQ85$ & $1000$ & $31$ & $57 \pm 42$ \\
W (Raw) & $KQ75$ & $1000$ & $31$ & $41 \pm 45$ \\
\hline
Q & $KQ75$ & $500$ & $30$ & $47 \pm 59$ \\
Q & $KQ85$ & $1000$ & $31$ & $33 \pm 42$ \\
Q & $KQ75$ & $1000$ & $31$ & $9 \pm 44$ \\
Q (Raw) & $KQ85$ & $1000$ & $31$ & $-64 \pm 42$ \\
Q (Raw) & $KQ75$ & $1000$ & $31$ & $-21 \pm 44$ \\
\hline
Combined V and W & $KQ75$ & $1000$ & $30$ & $76\pm 38$ \\
\hline
\end{tabular}
\end{center}
\caption{The estimated values for $f_{NL}$ together with the $1 \sigma $ error bars.}
\label{table:results}
\end{table}

As we see there is a large improvement in error bars from
$\ell_{max}=700$ to $\ell_{max}=1000$. This may seem at first sight surprising taking into account the fact that this range is noise dominated. However, this result is not unexpected if we
take into account the way the needlets 
are constructed. Indeed, $\ell_\mathrm{max}=700$ means that no needlet scale using 
information above $\ell_\mathrm{max}=700$ is included. 
Nevertheless, from the previous description of the needlet systems it is easy to see that 
the information from multipoles close to the boundary value $\ell_\mathrm{max}=700$ will
receive very little weight in general. Of course, the next needlet scale will contain 
information below as well as above $\ell_\mathrm{max}=700$. Therefore, when extending the 
analysis to higher $\ell$'s we do not only exploit $300$ more multipoles, but we are 
also able to extract better information from the multipoles below $\ell = 700$.

Another test was performed to take advantage of the fact that the CMB
should be independent of frequency, while the noise differs between
the channels. A data vector was composed from the needlet bispectrum
of both the individual V and W frequency channels.
\begin{equation}
  \mathbf{d} = \left[
    \begin{matrix}
      I^V_{j_1 j_2 j_3} \\
      \vdots \\
      I^W_{j_1 j_2 j_3} \\
      \vdots \\
    \end{matrix}
  \right]
\end{equation}
The full covariance matrix in this case contains information about
correlations between the frequencies, and should therefore enable us
to get smaller error-bars on $f_{NL}$.  However, for this analysis it
was necessary to use only $30$ needlet scales from each frequency
channel in order to get an invertible covariance matrix. And the
result (shown in the bottom row of table \ref{table:results}) was not
a large improvement from the analysis of the VW band at $31$ needlet
scales. However this is our estimate for $f_{NL}$ with the smallest
error-bars while using the $KQ75$ mask.

For the $B=1.2050$ case for the V+W band with the KQ75 mask, we have also investigated the change
in $f_{NL}$ as a function of the number of needlet scales included.
We thus included only the first 25 needlet scales (up to $\ell_\mathrm{max}=324$),
then the 26 first scales (up to $\ell_\mathrm{max}=390$ and so on up to all 31 scales.
The results are presented in table \ref{table:results2}. As expected, we see that the error bars
are decreasing with increasing $\ell_\mathrm{max}$. Differently from our case, in the optimal bispectrum estimation performed by the WMAP team and other groups error bars saturates earlier than lmax = 1000 because the full inverse covariance weighting scheme is not implemented and an approximation is used (whereas in this case the Monte Carlo approach used to estimate the bispectrum automatically accounts for this issue). Note however the WMAP error bars at lmax = 700 are still smaller than ours at lmax = 1000 because we don't implement a minimum variance estimator and thus we don't saturate the Rao-Cramer bound. Moreover an optimal bispectrum estimator with full inverse covariance weighting has been very recently implemented by Smith et al. 2008.

\begin{table}[htdp]
\begin{center}
\begin{tabular}{|l|c|c|c|c|c|}
\hline
$\ell_{max}$ & $n_j$ & V+W & V & W & Q \\
\hline
$324$ & $25$ & $64$ ($\pm 71$) & $77$ ($\pm 73$) & $63$ ($\pm 75$) & $26$ ($\pm 74$)\\
$390$ & $26$ & $44$ ($\pm 61$) & $81$ ($\pm 64$) & $35$ ($\pm 66$) & $25$ ($\pm 66$)\\
$471$ & $27$ & $44$ ($\pm 55$) & $71$ ($\pm 60$) & $40$ ($\pm 62$) & $31$ ($\pm 62$)\\
$567$ & $28$ & $42$ ($\pm 52$) & $56$ ($\pm 56$) & $55$ ($\pm 58$) & $43$ ($\pm 56$)\\
$683$ & $29$ & $73$ ($\pm 49$) & $71$ ($\pm 52$) & $72$ ($\pm 54$) & $23$ ($\pm 50$)\\
$823$ & $30$ & $81$ ($\pm 43$) & $80$ ($\pm 46$) & $72$ ($\pm 49$) & $24$ ($\pm 46$)\\
$1000$ & $31$ & $89$ ($\pm 39$) & $105$ ($\pm 42$) & $54$ ($\pm 45$) & $9$ ($\pm 44$)\\
\hline
\end{tabular}
\end{center}
\caption{The estimated values for $f_{NL}$ for different number $n_j$ of needlet scales. Since error bars
increase when using few needlet scales, the corresponding $1 \sigma$ error estimate is given in parenthesis.}
\label{table:results2}
\end{table}

\subsection{Consistency checks}
We see from these results that the estimates using the $KQ85$ mask
differs notably from the estimates using the $KQ75$ mask. This is
particularly the case for the $V+W$ channel, when only considering
scales up to $\ell_{max}=700$. This estimate when using the $KQ85$ mask
($f_{NL} = 156$) is much higher than the estimate found from
the same map, using the $KQ75$ mask ($f_{NL} = 88$). We are therefore
motivated to study simulations to find how often a change of mask
triggers such a large difference in the estimate.

We consider two sets of $10 000$ CMB sky simulations, each
set generated using the same random seed, and therefore identical. One set
is multiplied with the $KQ75$ mask, while the other is multiplied with
the $KQ85$ mask. Now we estimate $f_{NL}$ for both sets, and find the
difference between each estimate, and the corresponding estimate from
the identical map with the other mask, $\Delta f_{NL} = f_{NL}^{KQ75}
- f_{NL}^{KQ85}$. Then the mean value and standard deviation of
$\Delta f_{NL}$ is found.

For $\ell_\mathrm{max}=1000$ we found $\Delta f_{NL}=28$ for the WMAP data, whereas the
standard deviation $\sigma_{\Delta f_{NL}}=21$ for simulations. For $\ell_\mathrm{max}=700$,
we found $\Delta f_{NL}=68$ and $\sigma_{\Delta f_{NL}}=34$. In the first case, the shift
in $f_{NL}$ when changing mask is as expected whereas in the latter case,
the change $\Delta f_{NL}$ is slightly high, but only at the $2\sigma$ level.

As a further test of consistency, we also considered the difference in $f_{NL}$ estimate between
$\ell_{max} = 700$ and $\ell_{max} = 1000$ when using the $KQ85$ galactic cut $\Delta f_{NL}
=f_{NL}^{\ell_{max}=700} - f_{NL}^{\ell_{max}=1000}= 39$. However, a
comparison with simulations reveal that $\Delta f_{NL}$ have a
standard deviation of $\sigma = 30$. In other
words, the difference in the two estimates is well within $2 \sigma$
and is to be expected.

To test the variation of $f_{NL}$ with increasing galactic cut, we estimated $f_{NL}$ using 
the tiny Kp12 mask as well as the extended KQ75+ mask. We see that $f_{NL}$ decrease when 
going from the smallest mask to KQ85 and KQ75, but increases slightly again to KQ75+. 

At this point a $\chi^2$ test was implemented. Three identical sets of $10 000$ simulations 
were generated, and each set was multiplied with one of the KQ85, KQ75 and KQ75+ masks (we do not 
include the Kp12 mask as foregrounds are likely to be important for such a small mask).
For each simulation, a data vector, $\mathbf{d}$,
with two elements was formed from the difference in $f_{NL}$ estimates
between three different masks: 
\begin{equation}
\mathbf{d} = \left[
\begin{matrix}
f_{NL}^{KQ75} - f_{NL}^{KQ85} \\	
f_{NL}^{KQ75+} - f_{NL}^{KQ75} \\	
\end{matrix}.
\right]
\end{equation}

Of the $10000$ simulations, $5000$ were used to find a mean value and
covariance matrix for $\mathbf{d}$. Then a $\chi^2$ value was found
for each of the remaining $5000$ simulations as follows:
\begin{equation}
  \chi^2 = \left(\mathbf{d}-\langle \mathbf{d} \rangle\right)^TC^{-1}\left(\mathbf{d}-\langle \mathbf{d} \rangle\right) \label{eq:chi2}
\end{equation}
A similar $\chi^2$ value was found from the $f_{NL}$ values of the
WMAP data maps. Then the $\chi^2$ values for the simulations were
compared with that of the WMAP data. 

The result was that $37\%$ of the simulations had a higher value of $\chi^2$. We 
conclude that the variation of the $f_{NL}$ estimate for different masks (larger than Kp12) are 
within expectations for a Gaussian map.

As a further check for possible foreground contamination we will check the variation
of $f_{NL}$ with frequency channel. We investigated this by estimating $f_{NL}$
using $10 000$ simulated Gaussian CMB sky maps.  For each simulated
sky, three identical maps were generated. These maps were then
smoothed with the instrumental beam of the Q, V and W frequency
channel respectively, and noise was added independently to each of the
maps. For each of these sets of simulations a needlet transform was
performed using $31$ needlet scales in the range $2 \le \ell \le
1000$. Then the bispectra were found and $f_{NL}$ was estimated, using
between $25$ and $31$ of the needlet scales.

At this point we performed a $\chi^2$ test, similar to the one described above. First we tested the
variation between the frequency channels when using all $31$ of the
needlet scales. For every simulation, a data vector, $\mathbf{d}$,
with two elements was formed from the difference in $f_{NL}$ estimates
between the channels.

\begin{equation}
  \mathbf{d} = \left[
    \begin{matrix}
      f^Q_{NL} - f^V_{NL}  \\
      f^V_{NL} - f^W_{NL} \\
    \end{matrix}
  \right]
\end{equation}
The results using the KQ75 mask showed that only
$4.5\%$ of the simulations had a higher $\chi^2$ value, than the WMAP
data. This corresponds to a $\approx 2 \sigma$ deviation.

To investigate whether this is consistent on several scales, we also
performed the same test with some of the needlet scales removed.
This was done using between $25$ and $31$ needlet scales.
Finally we combined the data vectors from all these tests:
\begin{equation}
\mathbf{d} = \left[
\begin{matrix}
      f^{Q31}_{NL} - f^{V31}_{NL}  \\
      f^{V31}_{NL} - f^{W31}_{NL}  \\
      f^{Q30}_{NL} - f^{V30}_{NL}  \\
      f^{V30}_{NL} - f^{W30}_{NL}  \\
      \vdots \\
      f^{Q25}_{NL} - f^{V25}_{NL}  \\
      f^{V25}_{NL} - f^{W25}_{NL}  \\
\end{matrix}
\right]
\end{equation}
and used this to make a combined test. From the KQ75 results (shown in table
\ref{table:results5}) it seems that only by using all available scales
we find a small inconsistency of the $f_{NL}$ values between frequency channels.
We repeated the latter test using the smaller KQ85 cut and found in this case
that $14\%$ of the simulations had a higher $\chi^2$ concluding that foreground residuals
do not appear to be causing the difference in $f_{NL}$ for different channels.

\begin{table}[htdp]
\begin{center}
\begin{tabular}{|l|c|c|c|c|c|}
\hline
$\ell_{max}$ & needlet scales & $\%$ of sim. with higher $\chi^2$ \\
\hline
$324$ & $25$ & $20.6$  \\
$390$ & $26$ & $21.3$  \\
$471$ & $27$ & $54.6$  \\
$567$ & $28$ & $95.0$  \\
$683$ & $29$ & $43.2$  \\
$823$ & $30$ & $33.4$  \\
$992$ & $31$ & $4.5$  \\
\hline
\multicolumn{2}{|l|}{all of the above} & $12.2$ \\
\hline
\end{tabular}
\end{center}
\caption{Test of difference between frequency channels using different number of scales.
The last row combines all the other variables in one test.
  The table shows percentage of simulations with higher $\chi^2$ than the WMAP data using the KQ75 mask.}
\label{table:results5}
\end{table}

To test the influence of foregrounds on the estimate of $f_{NL}$, we
have estimated $f_{NL}$ on the WMAP maps before foreground subtraction
(raw maps). The results are listed in table \ref{table:results}. We
see in particular for the Q band that the the value of $f_{NL}$ is
negatively biased by the presence of foregrounds. A similar result was
also found in \citep{amit,komatsu:2008}. Foreground residuals would
thus be expected to give a too low value of $f_{NL}$. To check the
power of our consistency test, we repeated the above $\chi^2$ test of
the differences in estimated $f_{NL}$ between frequency channels using
31 scales. We find that only $0.7\%$ of the simulations have a higher
$\chi^2$ than for the WMAP data for the KQ75 cut, and none of the
simulations have a similarly high $\chi^2$ for the KQ85 cut. The test
thus shows a clear detection of foreground residuals in this case.

A similar $\chi^2$ test was now performed, but this time to study variation
between different number of needlet scales (and thus also different
$\ell_{max}$) used for the estimation:
\begin{equation}
  \mathbf{d} = \left[
    \begin{matrix}
      f^{31}_{NL} - f^{30}_{NL}  \\
      f^{30}_{NL} - f^{29}_{NL} \\
      \vdots \\
      f^{26}_{NL} - f^{25}_{NL} \\
    \end{matrix}
  \right]
\end{equation}
where the superscript denotes number of needlet scales used to
estimate the $f_{NL}$ value. $\chi^2$ was found using equation
(\ref{eq:chi2}) for the WMAP data as well as for the $5000$
simulations according to the same procedure as above. This was done
for the individual Q, V and W frequency channels, and for the combined
V+W map. The test was also performed using a combined data vector from
all the three frequency channels.  The results (table
\ref{table:results4}) show that the variation in the $f_{NL}$ estimate
with respect to needlet scales is well within the expected bounds.

\begin{table}[htdp]
  \begin{center}
    \begin{tabular}{|l|c|}
      \hline
      Freq. channel & $\%$ of sim. with higher $\chi^2$ \\
      \hline
      Q & $91.9$ \\
      V & $44.3$ \\
      W & $72.2$ \\
      V + W & $76.6$ \\
      combined Q, V and W& $56.3$\\
      \hline
    \end{tabular}
  \end{center}
  \caption{Test of $f_{NL}$ variation with respect to scale.
  Fraction of simulations with higher $\chi^2$ value than WMAP data.
  The results show that the WMAP data is consistent with Gaussianity in this respect.}
  \label{table:results4}
\end{table}

\section{Conclusions}
\label{sec:concl}
We have tested an estimator for $f_{NL}$ based on the needlet bispectrum \citep{needbisp}.
We used the estimator to obtain best fit values of $f_{NL}$ from the WMAP 5 year data,
using the combined V+W map as well as the independent frequency channels. The error bars on $f_{NL}$ obtained with the needlet bispectrum are significantly larger that those obtained by the optimal bispectrum estimator \citep{smith}, but the needlet bispectrum still provides an important and independent test of consistency.
We have further introduced a set of consistency tests based on the difference
 $\Delta f_{NL}=f_{NL}^1-f_{NL}^2$ where 1 and 2 refer to different masks,
 different frequency channels or different number of multipoles. We compare the
 differences $\Delta f_{NL}$ for the different cases to the values obtained in simulations.

We find our best estimate of $f_{NL}$ using the combined bispectrum
from the V and W channels giving $f_{NL}=76\pm38$ using the KQ75
mask and $\ell_\mathrm{max}=1000$, consistent within $1\sigma$ with the value of
$f_{NL}=51\pm32$ obtained by the WMAP team  as well as with the values obtained by \citep{amit,smith} all using
$\ell_\mathrm{max}=750$.

Using the combined $V+W$ map and $\ell_\mathrm{max}=1000$, we
obtained $f_{NL}=84\pm40$ for KQ75 and $f_{NL}=110\pm37$ using KQ85 (corrected for point source bias).
This difference in $f_{NL}$ using the two different masks was found
to be within the $2\sigma$ limit from simulations and thus
consistent with expectations. In order to further limit the risk of foreground contamination, we estimated $f_{NL}$ on an extended KQ75 mask excluding $37\%$ of the sky giving $f_{NL}=97\pm42$.

Using the independent frequency
channels and the KQ75 cut, we obtained $f_{NL}=9\pm44$ for the Q
band, $f_{NL}=105\pm42$ for the V band and $f_{NL}=54\pm45$ for the
W band. Such a large difference in $f_{NL}$ between the bands were
found only in $4.5\%$ of the simulations. This is $2\sigma$ away
from the expected value. This could be a sign of foreground
residuals but could also well be a statistical fluke. We found the
latter explanation to be more reasonable considering that for the
smaller KQ85 mask $14\%$ of the simulations had a larger difference.
Similar tests were made with values of $f_{NL}$ obtained using
different number of multipoles and channels, and no significant
deviations from the expected differences were found. We therefore
conclude that there are no convincing evidence of foreground
residuals having influenced the estimated value of $f_{NL}$, even
using the KQ85 galactic cut. However repeating these tests on the
next release of the WMAP data and on Planck data will be necessary
in order to confirm this claim.

\begin{acknowledgements}
FKH is grateful for an OYI grant from the Research Council of Norway.  This research has been partially supported by ASI contract 
I/016/07/0 "COFIS" and ASI contract Planck LFI Activity of Phase E2. We acknowledge the use of the NOTUR supercomputing facilities. We acknowledge the use of the HEALPix \citep{gorski:2005} package and the Legacy Archive for Microwave
  Background Data Analysis (LAMBDA). Support for LAMBDA is provided by
  the NASA Office of Space Science.
\end{acknowledgements}


\begin{thebibliography}{}
\bibitem[Acquaviva et al. (2003)]{standard1} Acquaviva, V., Bartolo, N., Matarrese, S. and Riotto, A. 2003, Nucl.\ Phys.\  B, 667, 119, arXiv:astro-ph/0209156

\bibitem[Alishahiha et al. (2004)]{dbi} Alishahiha, M., Silverstein, E. and Tong, D. 2004, Phys.\ Rev.\  D , 70, 123505

\bibitem[Arkani-Hamed et al. (2004)]{ghost} Arkani-Hamed, N., Creminelli, P., Mukohyama, S. and Zaldarriaga, M. 2004, JCAP, 0404, 001

\bibitem[Babich et al. (2004)]{creza} Babich, D., Creminelli, P., Zaldarriaga, M. 2004, Journal of Cosmology and Astroparticle Physics 8, 009

\bibitem[Baldi et al.(2006)]{bkmpAoS} Baldi, P., Kerkyacharian, G., Marinucci, D. and Picard, D. 2006, Annals of Statistics, Vol. 37, No. 3, 1150-1171, arxiv:math.st/0606599

\bibitem[Baldi et al. (2007)]{bkmpBer} Baldi, P., Kerkyacharian, G. Marinucci, D. and Picard, D. 2007, Bernoulli, Vol. 15, n.2, pp. 438-463, arxiv: 0706.4169

\bibitem[Baldi et al. (2008)]{denest} Baldi, P., Kerkyacharian, G., Marinucci, D., Picard, D. 2008, Annals of Statistics, in press, arxiv: 0807.5059

\bibitem[Bartolo et al. (2004a)]{bartoloreview} Bartolo, N., Komatsu, E., Matarrese, S. and Riotto, A. 2004a, Phys. Rept., 402, 103

\bibitem[Bartolo et al. (2004b)]{second1} Bartolo, N., Matarrese, S. and Riotto, A. 2004b, JCAP, 0401, 003

\bibitem[Bartolo et al. (2004c)]{second2} Bartolo, N., Matarrese, S. and Riotto, A. 2004c, JHEP, 0404, 006

\bibitem[Buchbinder et al. (2008)]{ek2} Buchbinder, E. I., Khoury, J. and Ovrut, B. A. 2008, Phys.\ Rev.\ Lett.\ , 100, 171302

\bibitem[Cabella et al. (2006)]{cabella06} Cabella, P., \ Hansen, F.K., Liguori, M., Marinucci, D., \ Matarrese, S., Moscardini, L., and Vittorio, N. 2006, MNRAS, 369, 819

\bibitem[Creminelli et al. (2006)]{creminelli} Creminelli, P. et al. 2006 JCAP, 5, 4
\bibitem[Cruz et al. (2007)]{Cruz1} Cruz, M., Cayon, L., Martinez-Gonzalez, E., Vielva, P. \& Jin, J. 2007, ApJ, 655, 11

\bibitem[Cruz et al. (2006)]{Cruz2} Cruz, M., Cayon, L., Martinez-Gonzalez, E., Vielva, P. 2006, MNRAS, 369, 57

\bibitem[Curto et al. (2008)]{curto} Curto, A. et al 2008, in press, arXiv:0807.0231

\bibitem[Delabrouille et al. (2008)]{dela08} Delabrouille, J., Cardoso, J.-F. , Le Jeune, M. , Betoule, M., Fay, G. \& Guilloux, F. 2008, arxiv 0807.0773


\bibitem[Dvali et al. (2004)]{gamma1} Dvali, G., Gruzinov, A. and Zaldarriaga, M. 2004, Phys. Rev. D, 69, 023505

\bibitem[Enqvist \& Sloth (2002)]{curvaton1} Enqvist, K. and Sloth, M. S. 2002,
Nucl. Phys., B626, 395

\bibitem[Fay et al. (2008a)]{fay08} Fay, G., Guilloux, F. , Betoule, M. , Cardoso, J.-F., Delabrouille, J. \& Le Jeune, M. 2008a, Phys. Rev. D, D78, 083013

\bibitem[Fay et al. (2008b)]{fg08} Fay, G. and Guilloux, F. 2008b, arxiv: 0807.2162

\bibitem[Geller \& Mayeli (2007a)]{gm2} Geller, D. and Mayeli, A. 2007a, ArXiv.0706.3642

\bibitem[Geller \& Mayeli (2007b)]{gm1} Geller, D. and Mayeli, A. 2007b, arxiv: 0709.2452
\bibitem[Geller \& Marinucci (2008)]{spin-mat} Geller, D. and Marinucci, D. 2008, arxiv: 0811.2935.

\bibitem[Geller et al. (2008)]{ghmkp08} Geller, D., Hansen, F.K., Marinucci, D., Kerkyacharian and Picard, D. 2008, Physical Review D, D78, 123533

\bibitem[G{\'o}rski et al. (2005)]{gorski:2005} G{\'o}rski, K.M.,
  E. Hivon, A.J. Banday, B.D. Wandelt, F.K. Hansen, M. Reinecke, and
  M. Bartelmann, HEALPix: A Framework for High-resolution
  Discretization and Fast Analysis of Data Distributed on the Sphere,
  \apj, 622, 759-771, 2005

\bibitem[Guilloux et al. (2007)]{guilloux} Guilloux, F., Fay, G., Cardoso, J.-F.2007, arxiv 0706.2598

\bibitem[Hu et al. (2001)]{Hu} Hu, W. 2001, Physical Review D, 64, 8, id.083005

\bibitem[Kolb et al. (2005)]{gamma2} Kolb, E. W., Riotto, A. and Vallinotto, A. 2005, Phys.\ Rev.\  D, 71, 043513

\bibitem[Kolb et al. (2006)]{gamma3} Kolb, E. W., Riotto, A. and Vallinotto, A. 2006, Phys.\ Rev.\  D, 73, 023522

\bibitem[Komatsu \& Spergel (2001)]{ks} Komatsu, E. and Spergel, D.N. 2001, Physycal Review D, 63, 063002, arXiv:astro-ph/0005036

\bibitem[Komatsu et al. (2005)]{ksw} Komatsu, E., Spergel, D. N. \& Wandelt, B. D. 2005, ApJ, 634, 14

\bibitem[Komatsu et al.(2008)]{komatsu:2008} Komatsu, E., et al.\
  2008, \apjs, in press, astro-ph/0803.0547v2

\bibitem[Koyama et al. (2007)]{ek1} Koyama, K., Mizuno, S., Vernizzi, F. and Wands, D. 2007, JCAP, 0711, 024

\bibitem[Lan \& Marinucci (2008a)]{needbisp} Lan, X. \& Marinucci, D. 2008a, Electronic Journal of Statistics, Vol. 2, pp.332-367, (2008), arXiv:0802.4020

\bibitem[Lan \& Marinucci (2008b)]{lan2} Lan, X., Marinucci, D. 2008b, arxiv: 0805.4154

\bibitem[Liguori et al. (2007)]{liguori:2007} Liguori, M. et al.\
  2007, \prd, 76, 105016

\bibitem[Lyth \& Wands (2002)] {curvaton2} D. Lyth and D. Wands 2002,
Phys. Lett. B, 524, 5

\bibitem[Maldacena (2003)] {standard2} Maldacena, J. M. 2003, JHEP, 0305, 013 (2003), arXiv:astro-ph/0210603

\bibitem[Marinucci (2006)]{m} Marinucci, D. 2006, The Annals
of Statistics 34, 1, arxiv;math/0502434

\bibitem[Marinucci et al. (2008)]{mpbb08} Marinucci, D., Pietrobon, D., Balbi, A., Baldi, P., Cabella, P., Kerkyacharian, G., Natoli, P., Picard, D., Vittorio, N., 2008, MNRAS, 383, 539

\bibitem[Martinez-Gonzalez et al. (2002)]{smhw} Martinez-Gonzalez, E. et al. 2002, MNRAS, 336, 22

\bibitem[Mayeli (2008)]{Mayeli} Mayeli, A. 2008, arxiv: 0806.3009

\bibitem[McEwen et al. (2006)]{mcewen2} McEwen J.D., Hobson M.P., Lasenby A.N., Mortlock, D.J. 2006, MNRAS, 371, L50

\bibitem[Moroi \& Takahashi (2001)]{curvaton3} Moroi, T. and Takahashi, T. 2001,
Phys. Lett. B, 522, 215, [Erratum-ibid. 539, 303 (2002)].

\bibitem[Narcowich et al. (2006a)]{npw1} Narcowich, F.J., Petrushev, P. and Ward, J.D. 2006a, SIAM Journal of Mathematical Analysis 38, 2, 574

\bibitem[Narcowich et al. (2006b)]{npw2} Narcowich, F.J., Petrushev, P. and Ward, J.D. 2006b, Journal of Functional Analysis 238, 2, 530

\bibitem[Pietrobon et al. (2006)]{pbm06} Pietrobon, D., Balbi, A., Marinucci, D. 2006, Physical Review D, 74, 043524

\bibitem[Pietrobon et al. (2008a)] {pietrobon} Pietrobon, D. et al. 2008a, arXiv:0812.2478
\bibitem[Pietrobon et al. (2008b)]{pietrobon08} Pietrobon, D., Amblard, A., Balbi, A., Cabella, P., Cooray, A. \& Marinucci, D. 2008b, Phys. Rev. D.,  D78, 103504 

\bibitem[Smith et al. (2008)]{smith} Smith, K. M., Senatore, L. \& Zaldarriaga, M. 2009, arxiv: 0901.2572

\bibitem[Vielva et al. (2004)]{vielva04} Vielva P., Martinez-Gonzalez E., Barreiro B., Sanz J. \& Cayon L. 2004, ApJ, 609, pp. 22

\bibitem[Yadav et al. (2007a)]{amit2} Yadav, A. S. et al 2007a, arXiv:0711.4933

\bibitem[Yadav et al. (2007b)]{amit3} Yadav, A. S., Komatsu, E. \& Wandelt B. D. 2007b, ApJ, 664, 680

\bibitem[Yadav \& Wandelt (2008)]{amit} Yadav, A. S. \& Wandelt, B. D. 2008, Phys. Rev. Lett, 100, 181301

\bibitem[Yadav et al. (2007)]{ykwlhm} Yadav, A.P.S., Komatsu, E., Wandelt, B.D. , Liguori, M., Hansen, F.K. and Matarrese, S. 2007, arXiv:0711.4933

\end{thebibliography}
\end{document}